# Mammographic Image Enhancement using Digital Image Processing Technique

Ardymulya Iswardani
Information System Department
Stmik Duta Bangsa
Surakarta, Indonesia
ardymulya@stmikdb.ac.id

Wahyu Hidayat
Informatics Department
Stmik Duta Bangsa
SURAKARTA, Indonesia
wahyu_hidayat@stmikdb.ac.id

*Abstract*— **PURPOSES:** this study aims to perform microcalsification detection by performing image enhancement in mammography image by using transformation of negative image and histogram equalization. **METHOD:** image mammography with .pgm format changed to. jpg format then processed into negative image result then processed again using histogram equalization. **RESULT:** the results of the image enhancement process using negative image techniques and equalization histograms are compared and validated with MSE and PSNR on each mammographic image. **CONCLUSION:** Image enhancement process on mammography image can be done, however there are only some image that have improved quality, this affected by threshold usage, which have important role to get better visualization on mammographic image.

## I. INTRODUCTION

This section described the motivation background of the studies as follows:

Difficulties in early identification of cancer cell existencies affected by it natural ability to multiply, survive, spread and hide for a certain time [1]. Mamographic screening is the best method for early identification. This method use X-ray to check the patient organs [2], [3]. Cancer can be identified from the presence of microcalsification, microcalsification is a major feature of cancer, however, false identification and unable to get important clues of microcalsification presence often occur [1]–[4].

Difficulties in recognizing the existence of microcalsification can be caused by many things, but one of them caused by the process of digitization [2]. This digitization process may cause degradation such as noisy and blurry. Image enhancement technique believed to produce image with better quality [5].

Therefore, this study aims to perform image enhancement in mammography image in recognizing microcalsification. Transformation to negative image and histogram equalization in this study used to process the original mammography image. At initial steps original mammographic image load to the application, then secondly process the image use as input to negative image techniques this technique suited when the dark region dominant in the image[6], final step histogram equalization used to redistributed the pixel value to get optimal value [2].

## II. LITERATURE REVIEW

This section described recent studies and basic image enhancement theory as follows:

### A. Recent studies

Microcalsification has characteristics such as normal tissue, to distinguish it required segmentation techniques [3], segmentation process is a technique that aims to distinguish observation areas visually. However, the visual quality of the image is influenced by the density of the observation object[2]. Many techniques have been proposed in recognizing microcalsification [1], [4]. Lots of method used, among them equalization histogram can be used to sharpness improvement [7], then negative image fits when the dark region become the dominant feature [6].

### B. Digital image

Digital image can be defined as a two dimensional function *f(x, y)*, where *x* and *y* are spatial coordinates and the amplitude of *f* in any coordinate pair *(x, y)* is called the gray level of the image at that point [6]. Digitized image as shown in Fig 1.

$$f(x,y) = \begin{bmatrix} f(0,0) & f(0,1) & \cdots & f(0,N-1) \\ f(1,0) & f(1,1) & \cdots & f(1,N-1) \\ \vdots & \vdots & \ddots & \vdots \\ f(M-1,0) & f(M-1,1) & \cdots & f(M-1,N-1) \end{bmatrix}$$

Fig 1 digitized image

### C. Negative image

The transformation of the original image into a negative image is required with conditions if the dark areas become the dominant [6]. Transformation to negative image:

$$Gray_{New} = 255 - Gray_{Old} \qquad (1)$$

This operation produced negative image [2]. $Gray_{New}$ obtain by subtracting $Gray_{Old}$ with value 255.

*D. Histogram equalization*

This technique will redistributed pixel value to obtain optimal result [8]

$$w = \frac{C_w T h}{n_x n_y} \quad (2)$$

Where:

- $w$ = histogram equalization
- $c_w$ = histogram cummulative
- $t_h$ = *threshold* (*default*: 256)
- $n_x - n_y$ = image dimension

### III. RESEARCH METODOLOGY

This section described steps involved in this studies as follow:

*A. proposed method*

This part described sistematically approach as shown in Fig 2.

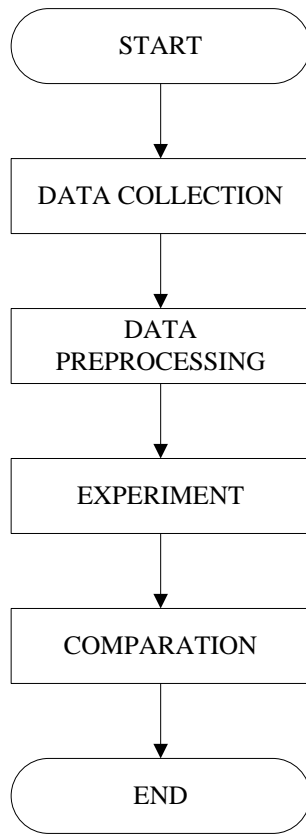

Fig 2 proposed method

- Data collection
  Mammographic image obtain from Mammographic Image Analysis Society (MIAS) from http://peipa.essex.ac.uk/info/mias.html. image group by normal and positive with cancer. Data format in .PGM (portable gray map). This studies required 8 image and group by normal and cancer positive.
- Data preprocessing
  Prepared the working directory for data saving, convert to jpg to make the image dimension same.
- Experiment
  .pgm format convert to jpg and transform to negative image and run histogram equalization.
- Comparation
  The final result of proposed method compared with the original image that already convert to .jpg format.

### IV. RESULT AND DISCUSSION

This section describe the result of the studies of image enhancement on mammographic image as follows:

*A. image group*

Image used in this studies obtain from Mammographic Image Analysis Society (MIAS) from http://peipa.essex.ac.uk/info/mias.html and group as shown Tab 1.

TABLE I. MAMMOGRAPHIC IMAGE

| CLASS | ABNORMALITY | CHAR | SAMPLE |
|---|---|---|---|
| NORM | | FATTY (F) | MDB006 |
| | | FATTY-GLANDULAR(G) | MDB007 |
| | | DENSE-GLANDULAR(D) | MDB003 |
| MALIGNANT | MICROCALCIFICATION | FATTY (F) | MDB231 |
| | | FATTY-GLANDULAR(G) | MDB209 |
| | | DENSE-GLANDULAR(D) | MDB239 |
| | WELL-DEFINED CIRCUMSCRIBED MASSES | FATTY (F) | MDB028 |
| | | FATTY-GLANDULAR(G) | MDB270 |

This table described mammographic image sample group by normal mammographic breast image and cancer breast image. Image enhancement applied to this eight sample image with fatty, fatty-glandular and dense-glandular.

*B. image processing*

Image format used in this studies is .pgm (portable gray map) with image dimension 1024 x 1024. When those image load in Octave the dimension change to 1200 x 898. There for next step is convert the .pgm image format to .jog image format. Format image transformation taken for MSE (Means Square Error) and PSNR (Peak Signal Noise Ratio) calculation. Different dimension makes the MSE and PSNR calculation failed.

*C. Experiment*

Image with .jpg format load to the application and used as input for negative process then process to histogram equalization. When two process done. The result compared with the image with .jpg format. The image enhancement process as show in Fig 3. Image processed to image negative develop by Integraged Development Environment GNU Octave, instruction to transformed the .jpg formatted to image negative shown as Fig 4 and instruction to run histogram equalization shown as Fig 5

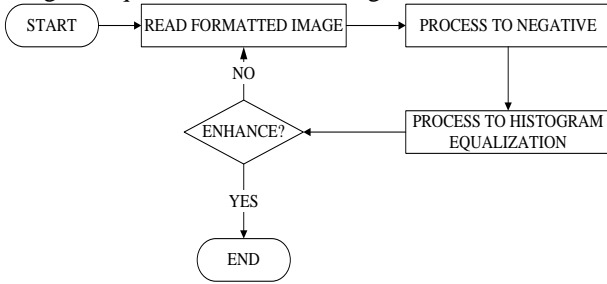

Fig 3 experiment

Fig 3 described the whole process in this studies. At initial steps, mammographic image with .pgm convert to .jpg format. This process need to be done since if the dimension of the image different the MSE and PSNR can't calculated.

```
1 function Negative (hObject, eventdata,
  NegCitra)
2     [fname, fpath] = uigetfile();
3     i = imread(fullfile(fpath,
  fname));
4     axes(NegCitra);
5     negatif = 255 - 1 - i;
6     imshow(negatif, []);
7  end
```

Fig 4 image negative instruction

Definition per line as follows:
Line 1 and 7  : user-defined function
Line 2        : dialog box *function*
Line 3        : read the data from line 2.
Line 4        : axes to display the image
Line 5        : image negative function
Line 6        : function to display the image

After the process of image negative done, continued to process this histogram equalization.

```
1 function  Histeq  (hObject, eventdata,
  Histeq)
2     [fname, fpath] = uigetfile();
3     i = imread(fullfile(fpath, fname));
4     negatif=255-1-i;
5     j = histeq(negatif, 256);
6     axes(Histeq);
7     imshow(j, []);
8 end
```

Fig 5 histeq function

Definition per line as follows:
Line 1 and 8  : user-defined function
Line 2        : dialog box
Line 3        : read data
Line 4        : image negative function
Line 5        : histeq threshold 256
Line 6        : axes
Line 7        : display the image

*D. MSE and PSNR*

The function of MSE (Means Square Error) and PSNR (Peak Signal Noise Ratio) is a common parameter used as an indicator in comparing the similarity of the two images (initial image and processing image). The use of both functions in this study basically aims as a measuring tool and / or to validate the level of similarity. The benefits of using these two functions as an alternative when encountering difficulties to finding experts in the field of image processing and cancer experts. Code to find the MSE and PSNR as shown in Fig 6:

```
1 img=imread();
2 img_result=imread();
3 [row, col, ~]=size(img);
4 mse = sum(sum((img-
  img_result).^2))/(row*col);
5 psnr = 10*log10(256*256/mse);
6 disp(mse);
7 disp(psnr);
```

Fig 6 MSE and PSNR

Definition per line as follows:
Line 1  :  read the image.
Line 2  :  read the result image
Line 3  :  array variable
Line 4  :  MSE.
Line 5  :  PSNR.
Line 6  :  display MSE.
Line 7  :  display PSNR.

MSE and PSNR show in Fig 7.

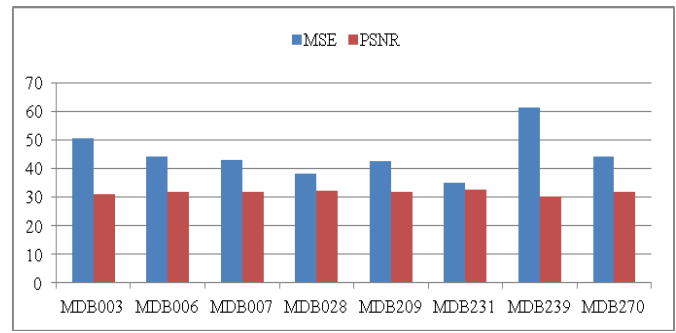

Fig 7 MSE and PSNR

Detail of MSE and PSNR value show in Tab. II. This table defined that only some of mammographic have better visualisation. however the overall process of image enhancement can applied to mammographic image.

TABLE II. MSE and PSNR

| IMAGE | MSE | PSNR |
|---|---|---|
| MDB003 | 50.35 | 31.145 |
| MDB006 | 43.99 | 31.731 |
| MDB007 | 42.9 | 31.84 |
| MDB028 | 37.99 | 32.368 |
| MDB209 | 42.6 | 31.871 |
| MDB231 | 35 | 32.725 |
| MDB239 | 61.44 | 30.281 |
| MDB270 | 44.13 | 31.718 |

*E. Comparation*

In this section will show the results of the use of image processing using image improvement techniques with the use of negative image function and histogram equalization. Both images are compared to be able to determine the image quality improvement. Improved imagery does not all have good quality images, but there are some image quality improvements. Histogram equalization threshold use 256 as default values. This quality improvement is used to facilitate the process of observation by health personnel. Result of the image enhancement process using negative image and histogram equalization show in Table III.

## V. CONCLUSION

Image enhancement process on mammography image can be done, however there are only some image that have improved quality, this affected by threshold usage, which have important role to get better visualization on mammographic image.

ACKNOWLEDGEMENT

THE RESEARCHER WOULD LIKE TO THANK THE DIRECTORATE OF RESEARCH AND COMMUNITY SERVICE OF DIRECTORATE GENERAL OF STRENGTHENING RESEARCH AND DEVELOPMENT OF RESEARCH, TECHNOLOGY AND HIGHER EDUCATION MINISTRY ACCORDING TO RESEARCH CONTRACT YEAR 2018.

TABLE III. RESULT COMPARATION

| MDB003_D_NORM | MDB003_D_NEG_NORM_HISTEQ | MDB006_F_NORM | MDB006_F_NORM_NEG_HISTEQ |
|---|---|---|---|
| 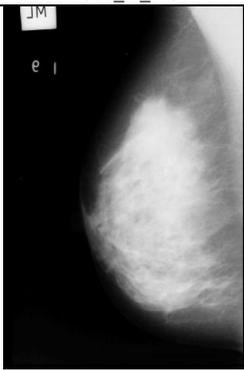 | 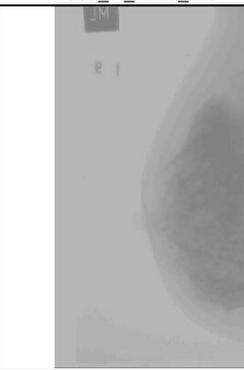 | 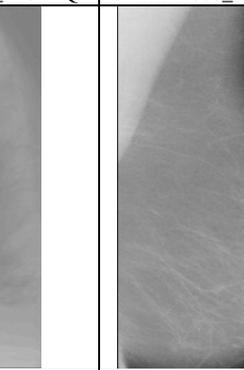 | 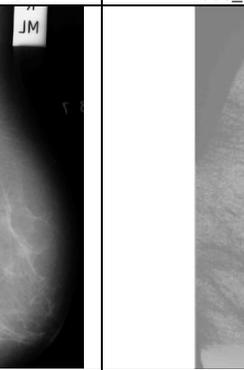 |
| MDB007_G_NORM | MDB007_G_NORM_NEG_HISTEQ | MDB028_F_CIRC | MDB028_F_CIRC_NEG_NORM_HISTEQ |
| 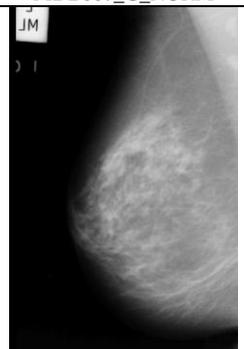 | 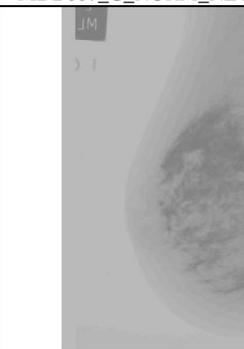 | 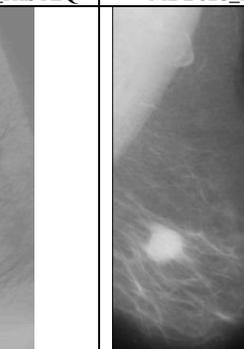 | 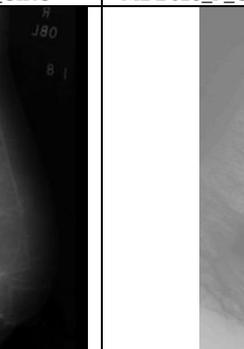 |

Continued...

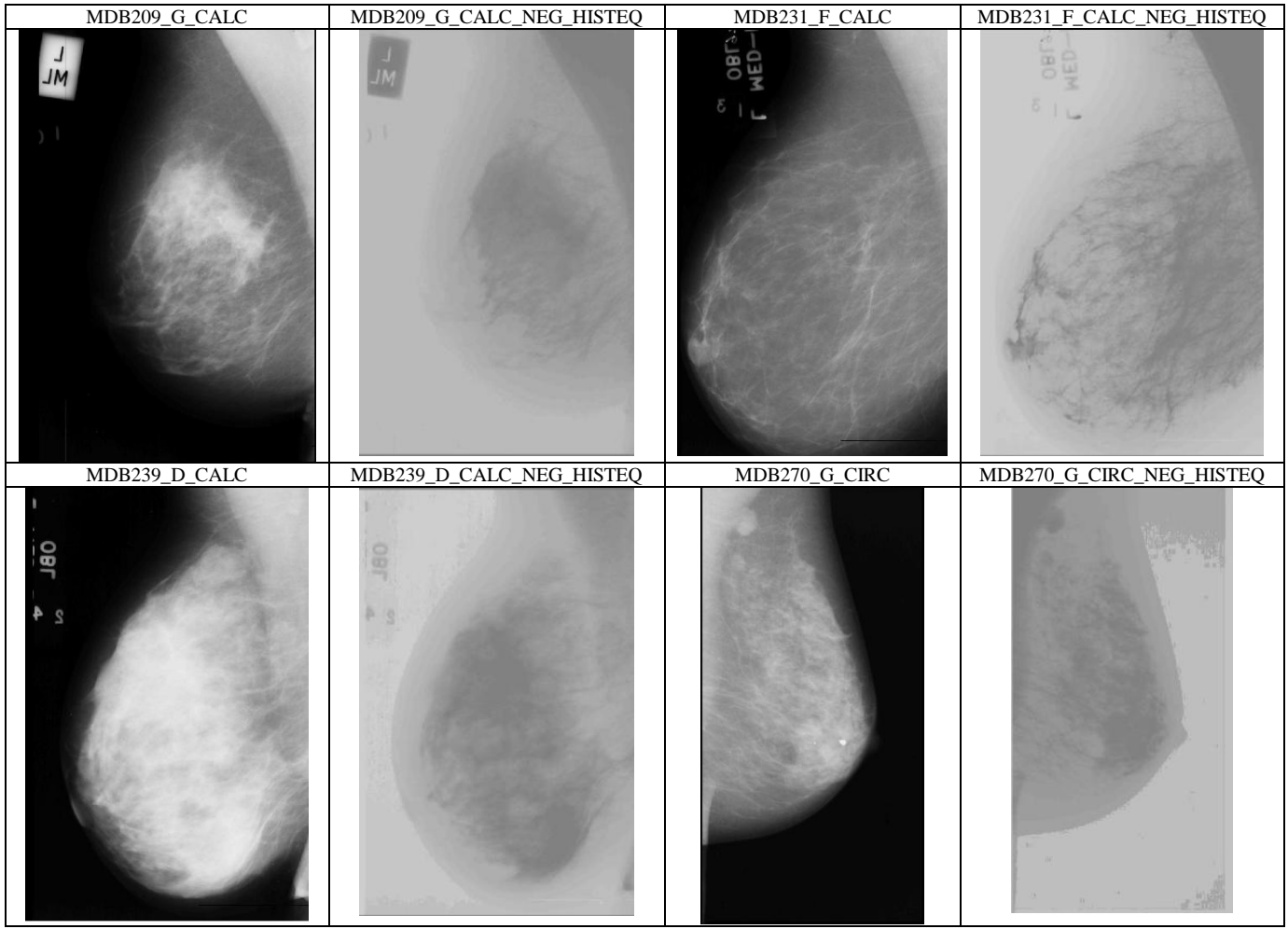